%% file: slepton.tex
\documentstyle[axodraw,graphicx,epsf]{aipproc2}
\begin{document}
\newcommand{\newc}{\newcommand}
\newc{\gev}{\,GeV}
\newc{\ra}{\rightarrow}
\newc{\rpv}{$\mathrm{\not\!R_p}$}
\newc{\beq}{\begin{equation}}
\newc{\eeq}{\end{equation}}
\newc{\ie}{{\it i.e.\/}\ }
\newc{\lam}{\lambda}
\newc{\ol}{\overline}
\title{Resonant Slepton Production}

\author{H. Dreiner$^*$, P. Richardson$^{\dagger}$\thanks{supported by
PPARC, UK.}, and M.H. Seymour$^*$}
\address{$^*$Rutherford Appleton Laboratory, Chilton, Didcot OX11 0QX, U.K.\\
$^{\dagger}$Department of Theoretical Physics,University of Oxford, 1
Keble Road, Oxford OX1 3NP, U.K.}

\maketitle

\begin{abstract}
  We consider the production of resonant sleptons via \rpv\
followed by gauge decays to a charged lepton and a neutralino which
then decays via \rpv. This gives a signature of two like-sign charged
leptons. We find a background at run II of $\mathrm{0.14\pm0.13}$ events with
an integrated luminosity of $\mathrm{2\,fb^{-1}}$. This enables us to probe
\rpv\ couplings of $\mathrm{2\times10^{-3}}$ for
slepton mass of $\mathrm{100\gev}$ and up to slepton masses of 
$\mathrm{300\gev}$
for \rpv\ couplings of $\mathrm{10^{-2}}$. 
\end{abstract}

\section*{Introduction}

Sleptons can be produced on resonance at the Tevatron via the \rpv\
$\mathrm{L_i Q_j{\ol D}_k}$ term in the superpotential. The slepton
can then decay again via the \rpv-Yukawa coupling. This has been
considered elsewhere \cite{hall,hewett,zerwas}. It can also decay via the
standard gauge decays of the MSSM \cite{hall} which we consider
here. There are two possible gauge decays of the charged slepton, \ie
\begin{eqnarray}
 \mathrm{\tilde{\ell}_{iL}^*} & \ra& \mathrm{\ell^{+}_{i} \tilde{\chi}^{0}},\\
 \mathrm{\tilde{\ell}_{iL}^*} & \ra& \mathrm{\bar{\nu}_{i} \tilde{\chi}^{+}}.
\end{eqnarray}
We focus on the decay of the slepton to a neutralino and charged lepton. The 
neutralino in turn will then decay via the same \rpv-operator 
$\mathrm{L_iQ_j{\ol D}_k}$,
\beq
\mathrm{\tilde{\chi}^{0}\ra \{\ell^-_i+u_j+ {\overline d}_k;\; \nu_i+d_j+ 
{\overline d}_k\}}.
\eeq
Since the neutralino is a Majorana fermion it can also decay to the charge
conjugate final states, with equal probability. In spirit, this is
similar to the HERA process considered in \cite{butterworth}.

\begin{figure}[t]
\begin{center} 
\input{feynman1}
\end{center}
\caption{Production of $\tilde{\chi}^{0}\ell^{+}$.\label{fig:cross}}
\begin{center} 
\input{feynman2}
\end{center}
\caption{LQD decay of $\mathrm{{\tilde\chi}^0}$. The neutralino is a Majorana
fermion and decays to the charge conjugate final state as well.
\label{fig:decay}}
\end{figure}

The tree-level Feynman diagrams for the slepton production and the
neutralino decay are shown in Fig.\,\ref{fig:cross} and Fig. 
\ref{fig:decay}, respectively. Due to the Majorana nature of the
neutralino, we have a signature of two like-sign charged leptons. In
the following we shall consider only electrons or muons, \ie we focus
on the operators $\mathrm{L_eQ_j{\ol D}_k}$ and $\mathrm{L_\mu 
Q_j{\overline D}_k}$.
We expect these leptons to have high transverse momentum, $p_{\mathrm{T}}$,
and be well isolated whereas the leptons from the Standard Model
backgrounds have lower $p_{\mathrm{T}}$ and are also poorly isolated. We
therefore hope that this signature can be seen above the background if
we apply isolation and $p_{\mathrm{T}}$ cuts.

\section*{Backgrounds}
In the following we combine the backgrounds for both electrons and
muons. The main backgrounds to this like-sign dilepton signature 
are as follows
\begin{enumerate}
\item \underline{$\mathrm{b\bar{b}}$ production} followed by the production of
at least one $\mathrm{B^{0}_{d,s}}$ meson, which undergoes mixing. If the two
b-quarks in the event decay semi-leptonically this gives two like-sign
charged leptons.
\item \underline{$\mathrm{t\bar{t}}$ production} followed by 
$\mathrm{t\ra W^{+}b\ra
e^{+} \bar{\nu_{e}} b}$, and $\mathrm{\bar{t}\ra W^{-}\bar{b}\ra q\bar{q}\bar
{b} \ra q \bar{q} W^{+}\bar{c}\ra q\bar{q} e^{+}\bar{\nu_{e}}\bar{c}}$.
\item \underline{Single top production} ($s$ and $t$ channel) followed by
semi-leptonic decays of the top and the B-meson produced after
hadronization.
\item \underline{Non-physics backgrounds} from fake leptons and charge
misidentification. There are also backgrounds due to the production of
weak boson pairs, \ie WZ and ZZ, where at least one of the charged
leptons is not detected \cite{nachtman}. These require a full
simulation including the detector. We do not consider them here.
\end{enumerate}

\begin{figure}[t]
\includegraphics[angle=90,width=0.45\textwidth,viewport=60 150 410 
550,clip]{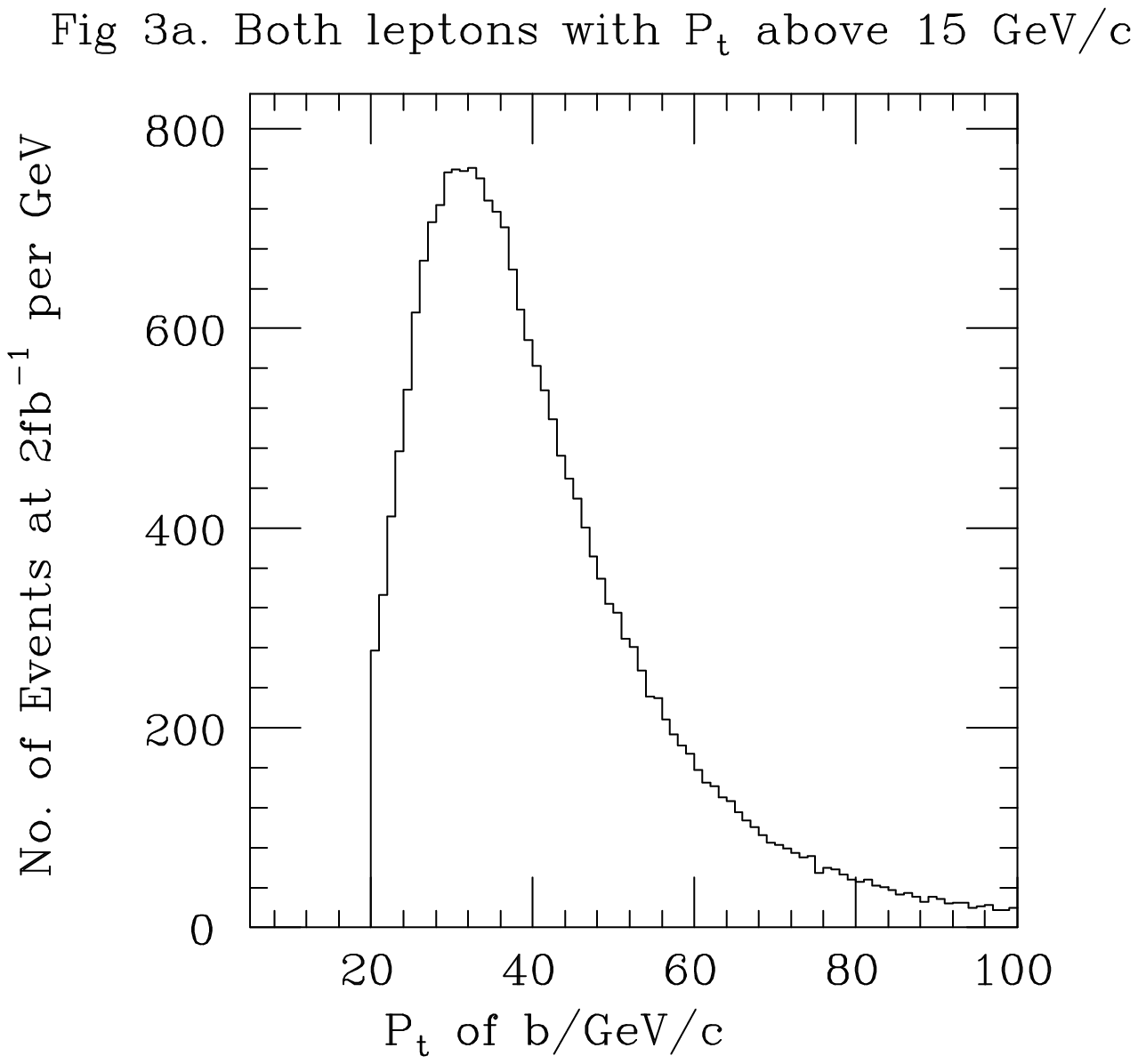}
\hfill
\includegraphics[angle=90,width=0.45\textwidth,viewport=60 150 410 
550,clip]{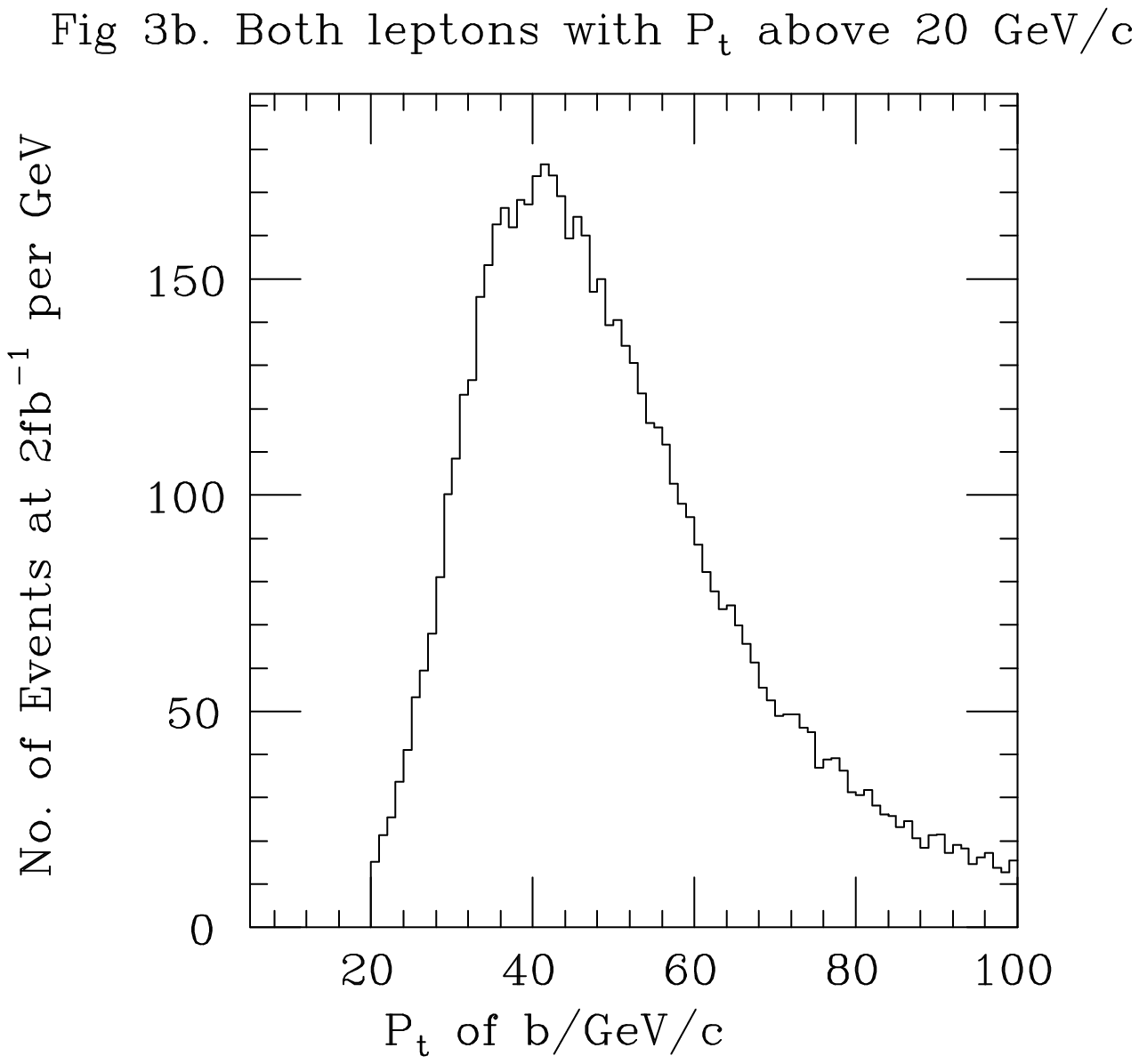}\\
\caption{Effect of the Parton-Level Cuts.\label{fig:parton}}
\end{figure}

We use HERWIG 6.0, \cite{HERWIGA,HERWIGB,HERWIGC}, to simulate these
background processes. The program includes the computation of the
supersymmetric spectrum and the MSSM decay branching ratios from the
ISASUSY program \cite{baertata}.  Due to the high cross section for
the production of $\mathrm{b \bar{b}}$ it was necessary to impose a
parton-level cut of $\mathrm{20\gev}$ on the $p_{\mathrm{T}}$ of the b
and $\mathrm{\bar{b}}$ to enable us to simulate a sufficient number of
events.  In Fig.  \ref{fig:parton} we show the distribution of events
(using the full Monte Carlo simulation) as a function of the
(parton-level) $p_{\mathrm{T}}$ of the bottom quark for two different
values of the lepton $p_{\mathrm{T}}$ cut. We did not simulate any
events for which the $p_{\mathrm{T}}$ of the bottom quark was below
$\mathrm{20\gev}$ since the cross section is too large. If we
extrapolate using Figs.\,\ref{fig:parton}a,\,b to lower b-quark
$p_{\mathrm{T}}$ we can see that for a lepton $p_{\mathrm{T}}$ cut of
$\mathrm{20\gev}$, Fig.\,\ref{fig:parton}b, our approximation should
be good, \ie we expect the area under the curve for $p_{\mathrm{T}}
\mathrm{(b)<20\gev}$ to be negligible. For $p_{\mathrm{T}}\mathrm{(
\ell)>15\gev}$, Fig.\,\ref{fig:parton}a, we would still expect a significant
number of events at $\mathrm{15\gev}<p_{\mathrm{T}}\mathrm{(b)<20\gev
}$. Besides the parton-level cut, we forced the B-mesons to decay
semi-leptonically.  This means we neglect the production of leptons
from the decay of charmed mesons which should also be a good
approximation as we expect the leptons produced from these decays to
be poorly isolated.

\begin{table*}[htp]
\caption{Summary of the Background Simulation\label{table1}}
\begin{tabular}{|l|ccc|}
Process & Cross section & No. of Events & Expected No. of Events after cuts, \\
& before Cuts /nb & simulated & for $\mathrm{2\,fb^{-1}}$ luminosity\\
\tableline
 $\mathrm{b\bar{b}}$ mixing & $(9.3\pm2.3)\times10^4$ & $8.3\times10^6$ &
 $0.12\pm0.12$ \\
 $\mathrm{t\bar{t}}$        & $6.81\pm0.31$           & $2.0\times10^5$ &
 $0.02\pm0.02$ \\
 single top                 & $1.55\pm0.12$           & $3.5\times10^4$ &
 $0.00\pm0.03$ \\
\tableline
 Total     & & & $0.14\pm0.13$ \\
\end{tabular}
\end{table*}

\begin{figure}[t]
\includegraphics[angle=90,width=0.45\textwidth,viewport=60 150 417 
550,clip]{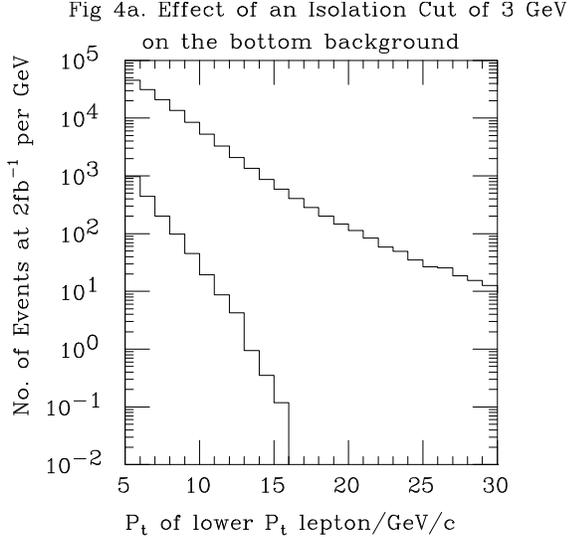}
\hfill
\includegraphics[angle=90,width=0.45\textwidth,viewport=60 150 417 
550,clip]{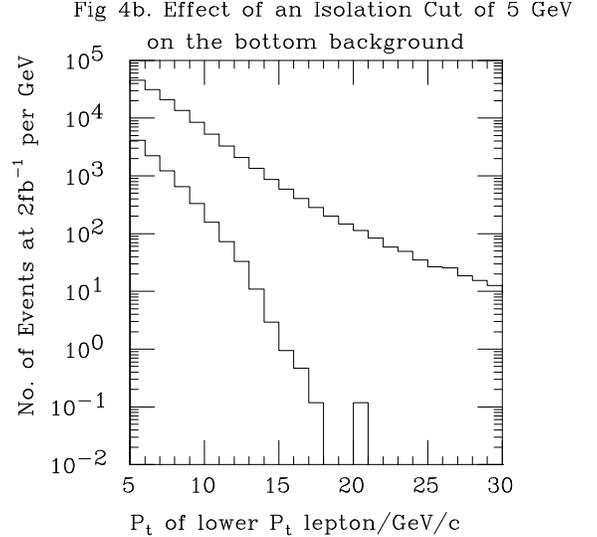}\\
\caption{Effect of the lepton isolation cut on the $\mathrm{b\bar{b}}$
background. The upper curve is the full $\mathrm{b\bar{b}}$ background and the
lower curve is obtained after imposing the isolation cut.
\label{fig:isocut}}
\end{figure}

\begin{figure}[t]
\includegraphics[angle=90,width=0.45\textwidth,viewport=60 150 417 
550,clip]{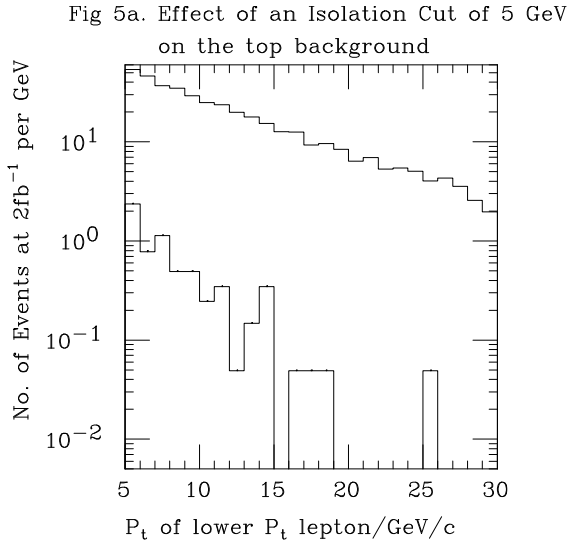}
\hfill
\includegraphics[angle=90,width=0.45\textwidth,viewport=60 150 417 
550,clip]{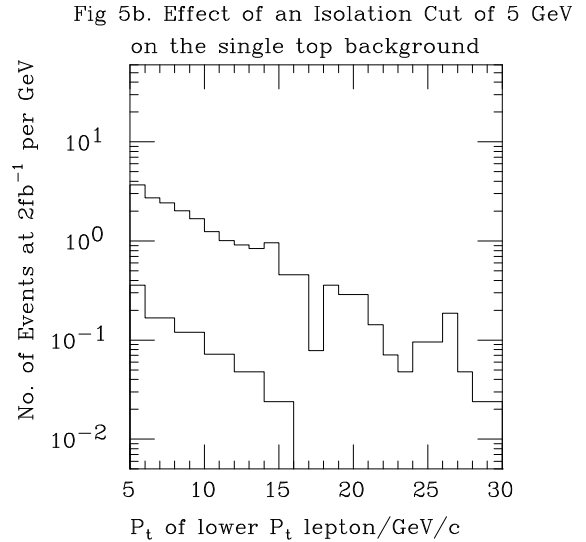}\\
\caption{Effect of the Isolation Cut on the $\mathrm{t\bar{t}}$ and single top 
Backgrounds\label{fig:isocut2}}
\end{figure}

Figure\,\ref{fig:isocut} displays the effect of the lepton isolation
cut on the $\mathrm{b\bar{b}}$ background for two different
values. The effect of the isolation cut on the $\mathrm{t\bar{t}}$ and
single top backgrounds is shown in Fig.\,\ref{fig:isocut2}. As can be
seen in Figs.\,\ref{fig:isocut},\,\ref{fig:isocut2}, by imposing an
isolation cut of $\mathrm{5\gev}$ and a cut on the $p_{\mathrm{T}}$ of
the leptons of $\mathrm{20\gev}$ the background can be almost
eliminated.  Table\,\ref{table1} shows the backgrounds with a
$p_{\mathrm{T}}$ cut on the leptons of $\mathrm{20\gev}$ and an
isolation cut of $\mathrm{5\gev}$. We have used the leading-order
cross section for the $\mathrm{b\bar{b}}$ and single top backgrounds
and the next-to-leading order cross section, with next-to-leading-log
resummation, from \cite{cross} for the $\mathrm{t\bar{t}}$ cross
section. In both cases the error on the cross section is the effect of
varying the scale between half and twice the hard scale, and the error
on the number of events is the error in the cross section and the
statistical error from the simulation added in
quadrature. Realistically we cannot reduce these statistical errors
due to the large number of events we would need to simulate. We have
implemented the full hadronization using HERWIG 6.0.

With these cuts and using Poisson statistics, a $5\sigma$ fluctuation
of the total background corresponds to 4 events with an integrated
luminosity of $\mathrm{2\,fb^{-1}}$. Hence we consider 4 signal events to be
sufficient for a discovery of the new \rpv\ signal process.

\section*{Signal}

To simulate the signal and the effect of the cuts, we modified HERWIG
6.0, \cite{HERWIGA,HERWIGB,HERWIGC}, to include the production
process, the MSSM decay of the slepton, and the \rpv\ decay of the
neutralino. The decay rate of the neutralino and its branching ratios
were calculated in the code and a matrix element for the neutralino
decay \cite{gondolo,morawitz} was implemented in the Monte Carlo simulation.

\begin{figure}[t]
\includegraphics[angle=90,width=0.45\textwidth,viewport=50 70 550
700,clip]{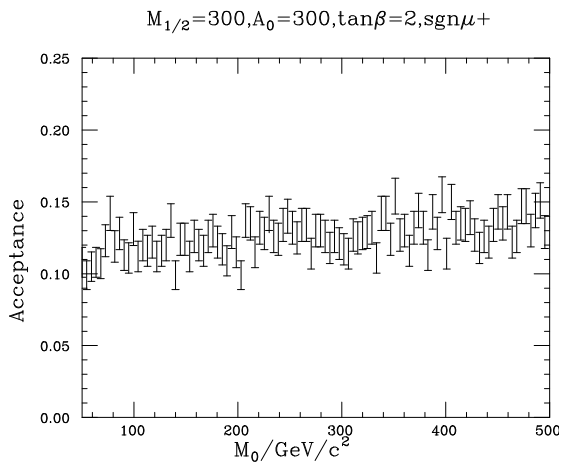}
\hfill
\includegraphics[angle=90,width=0.45\textwidth,viewport=50 70 550 
700,clip]{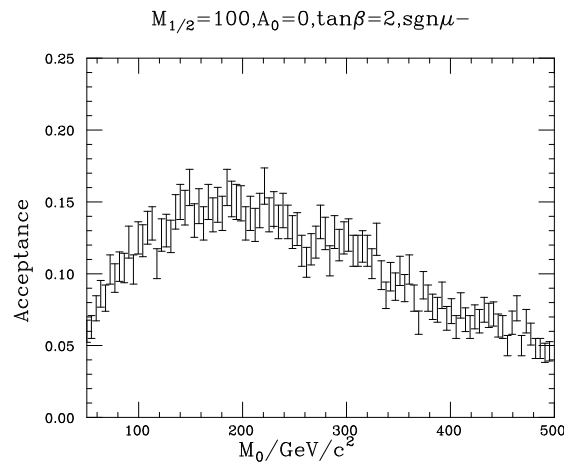}\\
\caption{Acceptance from Monte Carlo Simulation for two different
SUGRA points.\label{fig:eff}}
\end{figure}

We use the program to estimate the acceptance of the signal process,
\ie the fraction of the like-sign dilepton events which pass the cuts
multiplied by the branching ratio to give a like-sign dilepton event.
Fig.\,\ref{fig:eff} shows the acceptance for two different SUGRA
points, with an isolation cut on the leptons of $\mathrm{5\gev}$ and a cut 
$p_{\mathrm{T}}\mathrm{
(\ell)>20\gev}$. As can be seen in Fig.\,\ref{fig:eff}b, the acceptance
drops in two regions. For lower values of $M_0$, the slepton is not
much heavier than the neutralino. The charged lepton from the decay of
the slepton is then quite soft and gets rejected by the $p_{\mathrm{T}}$ cut. 
For
large values of $M_0$ the slepton is much heavier than the
neutralino. The neutralino then gets a significant boost from the
slepton decay. The neutralino decay products are folded forward in
the direction of this boost causing the event to be rejected by the
lepton isolation cut.

\begin{figure}[t]
\includegraphics[angle=90,width=0.45\textwidth,viewport=50 70 550 
720,clip]{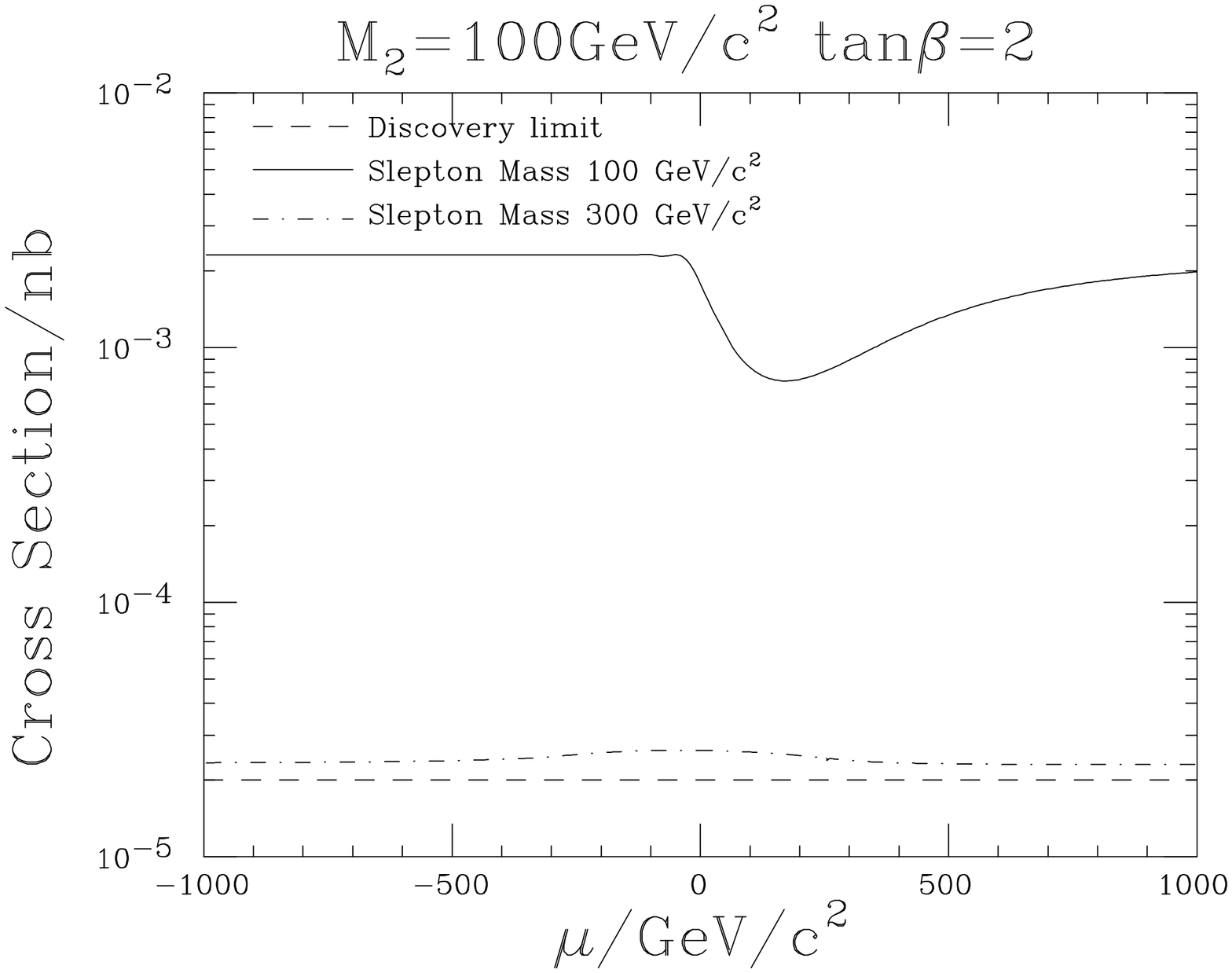}
\hfill
\includegraphics[angle=90,width=0.45\textwidth,viewport=50 70 550 
720,clip]{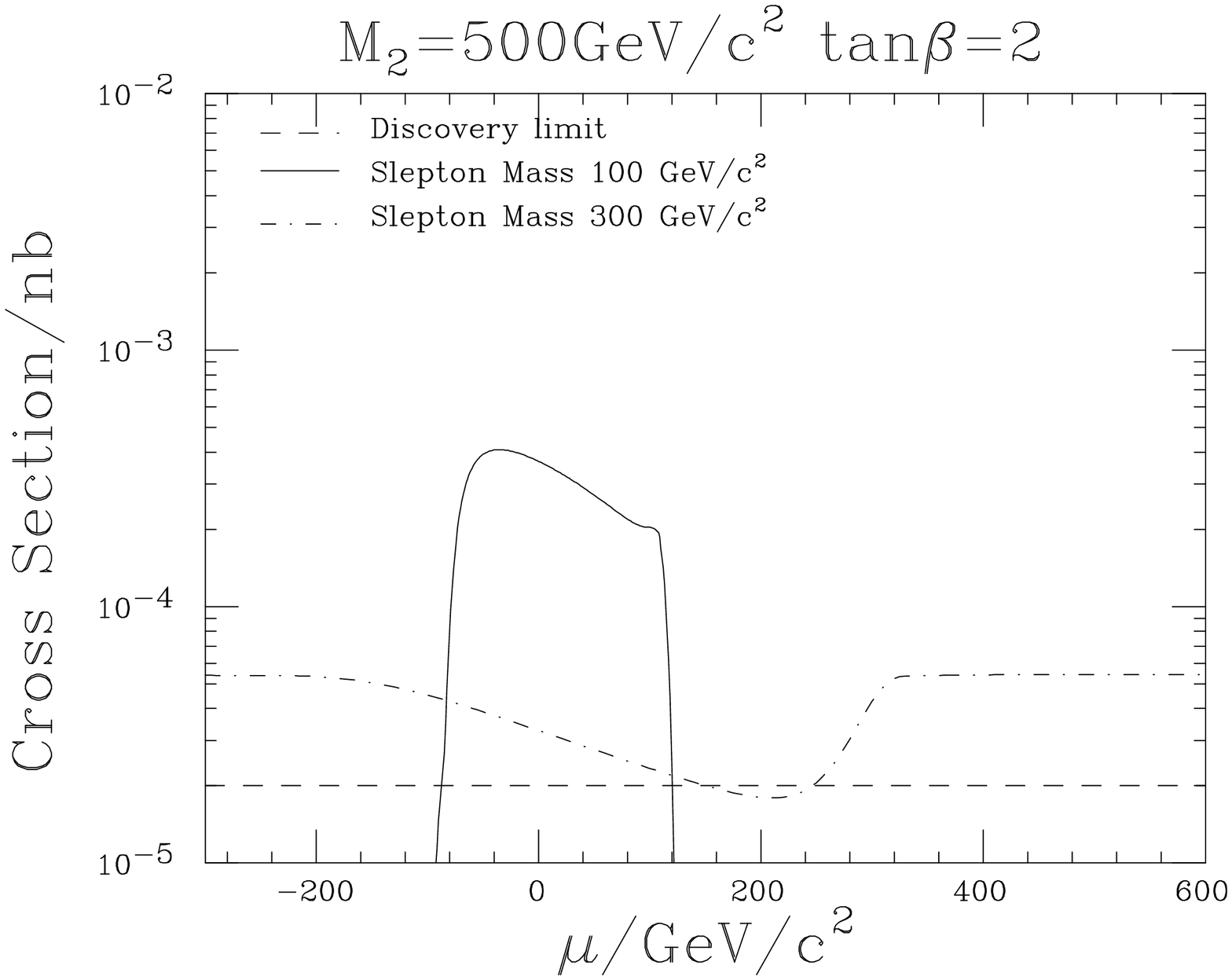}\\
\caption{Cross section for $\tan\beta=2$ and $\lam'_{211}=10^{-2}$. The
long-dashed horizontal line corresponds to the discovery limit of 4
signal events for $\mathrm{2\,fb^{-1}}$. The signal cross section after cuts is
given by the solid and dot-dashed curves for two slepton masses.
\label{fig:Xsect1}}
\end{figure}

To estimate the acceptance properly we need to run a scan of the
SUGRA parameter space using the Monte Carlo event generator. This
still remains to be done. To give some idea of what range of couplings
and masses we may be able to probe instead we assume an acceptance of
10\% using the same cuts as before. We can then estimate the range of
couplings which may be accessible.

\begin{figure}
\includegraphics[angle=90,width=0.45\textwidth,viewport=50 70 550 
720,clip]{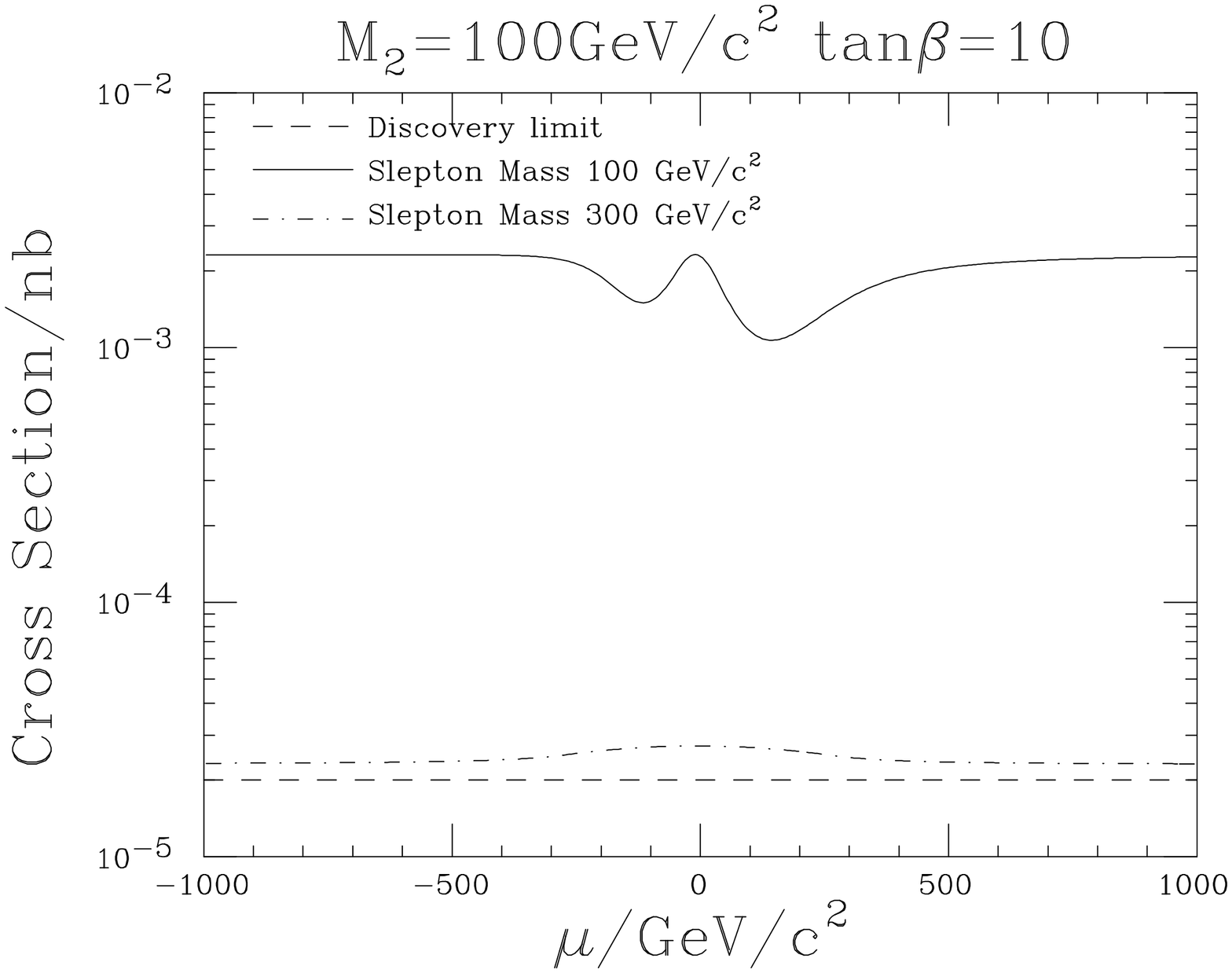}
\hfill
\includegraphics[angle=90,width=0.45\textwidth,viewport=50 70 550
 720,clip]{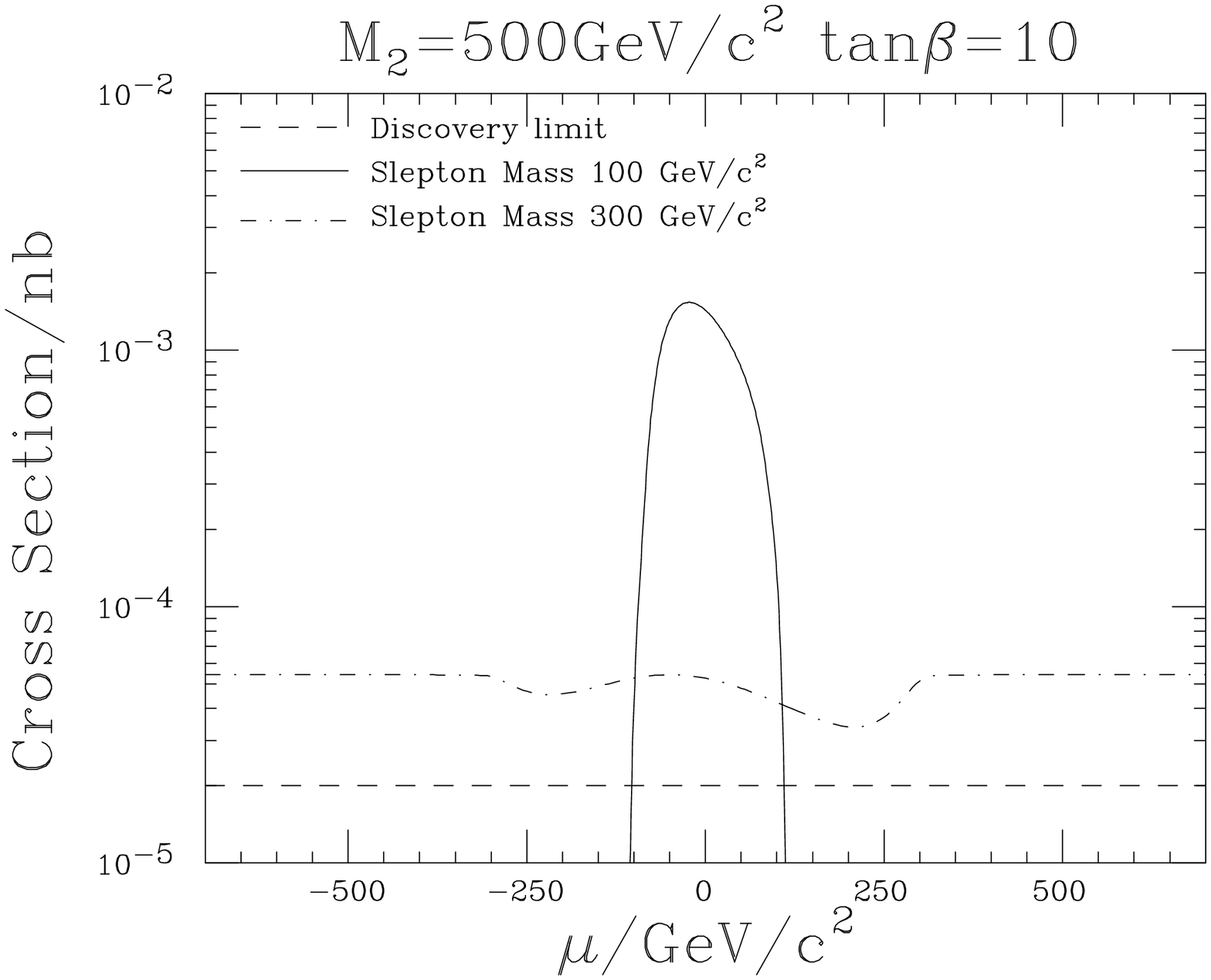}\\
\caption{Cross section for $\tan\beta=10$ and 
$\lam'_{211}=10^{-2}$.\label{fig:Xsect2}}
\end{figure}

As can be seen in Figs.\,\ref{fig:Xsect1},\,\ref{fig:Xsect2} the
production cross section for $\lam'_{211}=10^{-2}$ is sufficient to
produce a signal which is more than $5\sigma$ above the background for
large regions of the SUGRA parameter space. In some regions where the
neutralinos become Higgsino-like the cross section drops. The
cross section also drops as we approach the region where the
neutralino is heavier than the slepton and the resonance becomes
inaccessible.

We focused on the coupling $\lam'_{211}$ because the experimental
bound on $\lam'_{111}$ from neutrinoless double beta decay is very
strict \cite{klapdor,Herbi}. The bound on $\lam'_{111}$ weakens as the
squark mass squared and for squark masses above about $\mathrm{300\gev}$ (which
we expect in the SUGRA scenario for the heavier slepton masses)
$\lam'_{111}\approx10^{-2}$ is experimentally allowed and our analysis
thus applies for this case as well. $\lam'_{211}
\approx10^{-2}$ is well within the present experimental bounds \cite{Herbi}.

In these figures we see that we are sensitive to slepton masses up
to $\mathrm{300\gev}$ for couplings of $10^{-2}$. The production cross section
scales with the square of the coupling. For slepton masses around
$\mathrm{100\gev}$, just above the LEP limits, we can thus probe couplings down
to about $2 \times 10^{-3}$.

\section*{Conclusion}

We have performed an analysis of the physics background for like-sign
dilepton production at run II and find that with an integrated
luminosity of $\mathrm{2\,fb^{-1}}$, a cut on the transverse momentum
of the leptons of $\mathrm{20\gev}$ and an isolation cut of
$\mathrm{5\gev}$ the background is $0.14 \pm 0.13$ events. This means
that 4 signal events would correspond to a $5\sigma$ discovery,
although in a full experimental analysis the non-physics backgrounds
must also be considered.

Using a full Monte Carlo simulation of the signal including a
calculation of the neutralino decay rate, its partial widths and a
matrix element in the simulation of the decay we found that the
acceptance for the signal varies but for a reasonable range of
parameter space is 10\% or greater.

When we then look at the cross section for the production of
$\tilde{\chi}^{0}\ell^{+}$ we find that we can probe \rpv\ couplings
of $2\times10^{-3}$ for slepton mass of $\mathrm{100\gev}$ and up to
slepton masses of $\mathrm{300\gev}$ for \rpv\ couplings of $10^{-2}$,
and higher masses if the coupling is larger.

\end{document}

%% file: feynman1.tex
\begin{picture}(360,60)(0,30)
\SetScale{0.7}
\ArrowLine(5,76)(60,102)
\ArrowLine(60,102)(5,128)
\DashArrowLine(90,102)(60,102){5}
\ArrowLine(90,102)(145,76)
\ArrowLine(145,128)(90,102)
\Text(80,55)[]{$\mathrm{\tilde{\chi}^{0}}$}
\Text(80,90)[]{$\mathrm{\ell^{+}}$}
\Text(25,90)[]{$\mathrm{\bar{d}}$}
\Text(25,57)[]{$\mathrm{u}$}
\Text(53,78)[]{$\mathrm{\tilde{\ell}_{L}}$}
\Vertex(60,102){1}
\Vertex(90,102){1}
\SetScale{0.7}
\ArrowLine(240,128)(185,128)
\ArrowLine(240,128)(295,128)
\ArrowLine(240,76)(185,76)
\ArrowLine(295,76)(240,76)
\DashArrowLine(240,76)(240,128){5}
\Text(190,98)[]{$\mathrm{\tilde{\chi}^{0}}$}
\Text(190,47)[]{$\mathrm{\ell^{+}}$}
\Text(153,47)[]{$\mathrm{\bar{d}}$}
\Text(153,95)[]{$\mathrm{u}$}
\Text(160,70)[]{$\mathrm{\tilde{d}_{R}}$}
\Vertex(240,128){1}
\Vertex(240,76){1}
\ArrowLine(420,128)(365,128)
\ArrowLine(475,128)(420,128)
\ArrowLine(365,76)(420,76)
\ArrowLine(420,76)(475,76)
\DashArrowLine(420,76)(420,128){5}
\Text(315,47)[]{$\mathrm{\tilde{\chi}^{0}}$}
\Text(315,98)[]{$\mathrm{\ell^{+}}$}
\Text(277,98)[]{$\mathrm{\bar{d}}$}
\Text(277,47)[]{$\mathrm{u}$}
\Text(287,70)[]{$\mathrm{\tilde{u}_{L}}$}
\Vertex(420,128){1}
\Vertex(420,76){1}
\end{picture}

%% file: feynman2.tex
\begin{picture}(360,80)(0,0)
\SetScale{0.7}
\ArrowLine(5,78)(60,78)
\ArrowLine(105,105)(60,78)
\ArrowLine(84,53)(129,26)
\ArrowLine(129,80)(84,53)
\DashArrowLine(60,78)(84,53){5}
\Text(25,63)[]{$\mathrm{\tilde{\chi^{0}}}$}
\Text(55,70)[]{$\mathrm{\ell^{+}}$}
\Text(75,20)[]{$\mathrm{d}$}
\Text(75,54)[]{$\mathrm{\bar{u}}$}
\Text(45,40)[]{$\mathrm{\tilde{\ell}_{L}}$}
\Vertex(60,78){1}
\Vertex(84,53){1}
\ArrowLine(185,78)(240,78)
\ArrowLine(285,105)(240,78)
\ArrowLine(264,53)(309,26)
\ArrowLine(309,80)(264,53)
\DashArrowLine(240,78)(264,53){5}
\Text(150,63)[]{$\mathrm{\tilde{\chi}^{0}}$}
\Text(200,54)[]{$\mathrm{\ell^{+}}$}
\Text(200,20)[]{$\mathrm{d}$}
\Text(180,70)[]{$\mathrm{\bar{u}}$}
\Text(170,40)[]{$\mathrm{\tilde{u}_{L}}$}
\Vertex(240,78){1}
\Vertex(264,53){1}
\ArrowLine(365,78)(420,78)
\ArrowLine(420,78)(465,105)
\ArrowLine(489,26)(444,53)
\ArrowLine(489,80)(444,53)
\DashArrowLine(444,53)(420,78){5}
\Text(277,63)[]{$\mathrm{\tilde{\chi}^{0}}$}
\Text(330,20)[]{$\mathrm{\ell^{+}}$}
\Text(310,70)[]{$\mathrm{d}$}
\Text(330,54)[]{$\mathrm{\bar{u}}$}
\Text(300,40)[]{$\mathrm{\tilde{d}_{R}}$}
\Vertex(420,78){1}
\Vertex(444,53){1}
\end{picture}